# Is the Atmosphere Really Such a Good Blanket?

Slavoljub Mijovic

University of Montenegro; Faculty of Natural Sciences and Mathematics; Cetinjski put 2

20000 Podgorica MONTENEGRO e-mail: slavom@ucg.ac.me

**Abstract:** Wrongly used the current methodology for treating and averaging Earth surface temperatures made many contradictions and missintepretation in different scientific disciplines, particularly in the science of climate change and global warming. The methodology proposed earlier by this author eliminates the shortcomings and gives clear picture how to proceed further.

This paper demonstrates that the influence of the atmosphere on the warming of the Earth is highly overestimate, maybe at least ten degrees Celsius. The analysis is based on the ERA5 reanalysis dataset produced by the European Centre for Medium-Range Weather Forecasts (ECMWF), using the skin temperature data. Hourly data covering the period 1950–2025 were used on a global grid consisting of approximately 1.04 million cells representing the Earth's entire surface.

The proposed method allows absolute estimation of global warming and confirms the long-term warming trend and provides a comparative analysis with the results obtained using the conventional method.

**Key words:** climate change; greenhouse effect; atmosphere; ERA5; the new method, effective mean temperature.

## Introduction

The climate system primarily consists of the land, ocean, ice surfaces, atmosphere, and solar radiation that provides energy. These components interact to produce the conditions on and around the Earth's surface that we refer to as climate [1]. Climate is defined by averaging physical quantities that characterize these conditions over space and time, typically over a 30-year period [2]. The word “averaging” is common in the climate science but the key phrase is “How to average?”.

The proper method of averaging the Earth temperature is proposed in [3,4] . But, from the consequences of the widely accepted conventional method, how to estimate the Earth mean temperature, one can find in literature such statement [1] p. 104 “2. The background or pre-industrial levels of greenhouse gases warm the surface by about 33K and without this ‘natural’ greenhouse effect the Earth would be frozen everywhere.” Similar statements one can find elsewhere. It results from incorrect premises and the purpose of this work is to estimate the more realistic the contribution of the atmosphere in the Earth heating.

## THE METHOD

The basic idea of the new method is to relate the energy flux by which a celestial body cools to a corresponding temperature, previously termed the *effective temperature of potential cooling*. Since all celestial bodies ultimately cool exclusively by radiation according to the Stefan–Boltzmann law, the cooling energy flux is proportional to the fourth power of this effective temperature.

If Earth’s surface is divided into $N$ grid cells, with area $\Delta S_i$ and temperature $T_{i,j}$ of the $i-th$ cell during a time interval $\Delta t_j$, and if the total surface area of Earth is $S_{Earth}=\sum_{i=1}^{N}\Delta S_i$, then the energy flux of Earth’s potential cooling (in the absence of an atmosphere) is calculated as

$$\sum_{i=1}^{N}\Delta S_i\left(\sum_{j=1}^{M}\ \varepsilon_{i,j}\sigma T_{i,j}^{4}\Delta t_j\right), \tag{1}$$

where $\varepsilon_{i,j}$ −is the emissivity of the $i-th$ cell during the time interal $\Delta t_j$, $M$ is the total number of time intervals for the selected period $\tau=\sum_{j=1}^{M}\Delta t_j$ and $\sigma$ is the Stefan–Boltzmann constant.

To test how this energy flux, which is proportional to global warming, changed with time, ERA5 data from the European Centre for Medium-Range Weather Forecasts (ECMWF) were used, specifically the dataset *ERA5 hourly data on single levels from 1940 to present*. The variable selected was ERA5 skin temperature. This is the temperature of the thin surface layer that emits infrared radiation to space, making it the appropriate temperature field for calculating surface-emitted energy and effective radiative temperature. Hourly data for the period 1950–2025 were downloaded for approximately 1.04 million grid cells covering the entire globe.

The skin temperature was chosen because it allows the absolute estimation of global warming changes over considered time by calculating above sum, taking into account temporal and spatial changes of the surface emissivity (ocean, land, ice/snow).

With the assistance of artificial intelligence (ChatGPT, https://chatgpt.com/), a MATLAB script was developed to retrieve ERA5 data and compute the above sum, i.e., the total radiated energy $\boldsymbol{E_{tot}}$ for each year.

The corresponding effective temperatures were calculated as

$$T_{eff} = \left(\frac{E_{tot}}{\sigma S_{Earth}\tau}\right)^{1/4}. \tag{2}$$

$Teff$ is the temperature of a blackbody that would emit the same total energy as the actual Earth's surface, whose temperatures and emissivity vary spatially and temporally.

## RESULTS AND DISCUSSION

The results of the calculation are shown in **Figure 1**, **2, 3** and **4.**

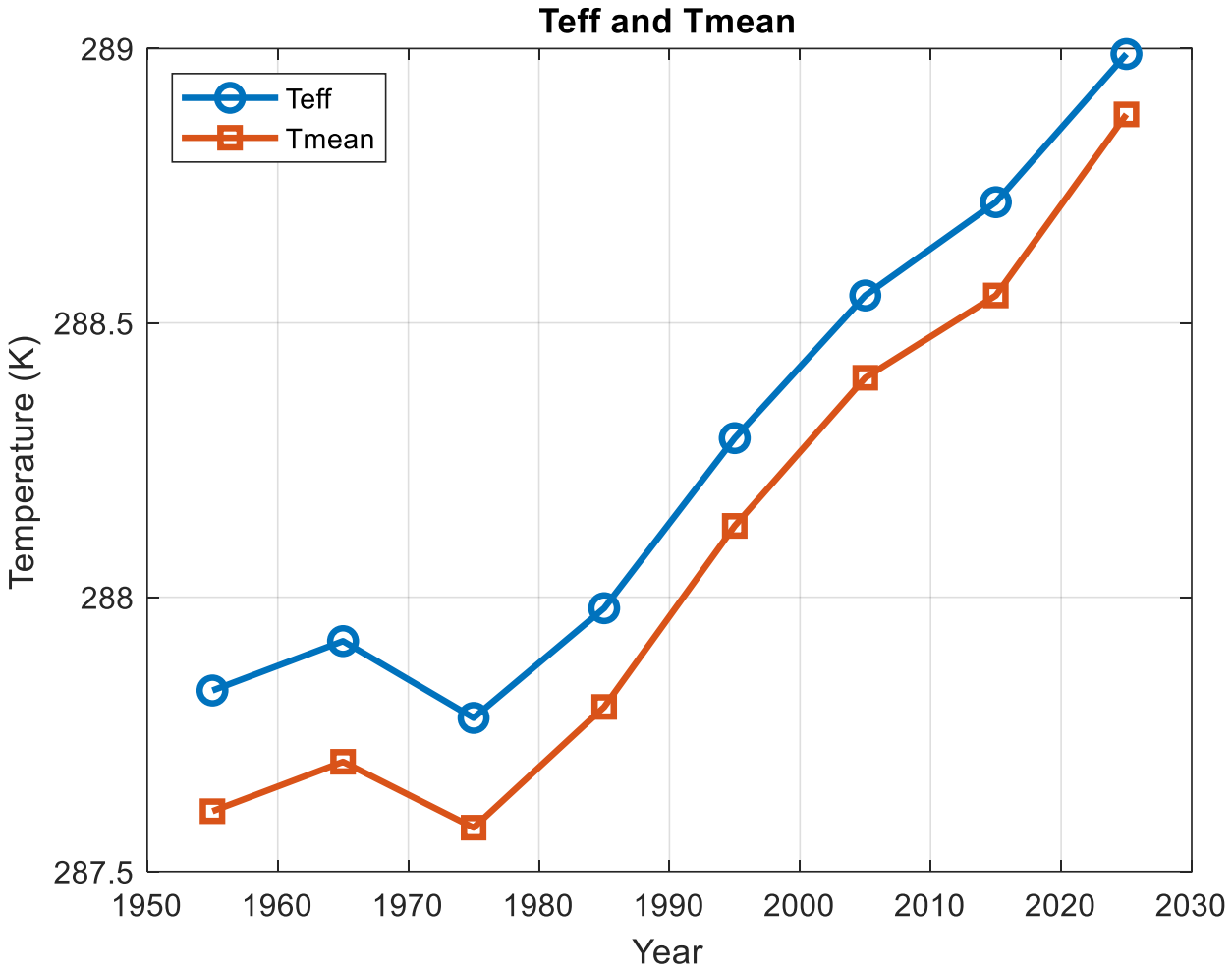

**Figure 1.**

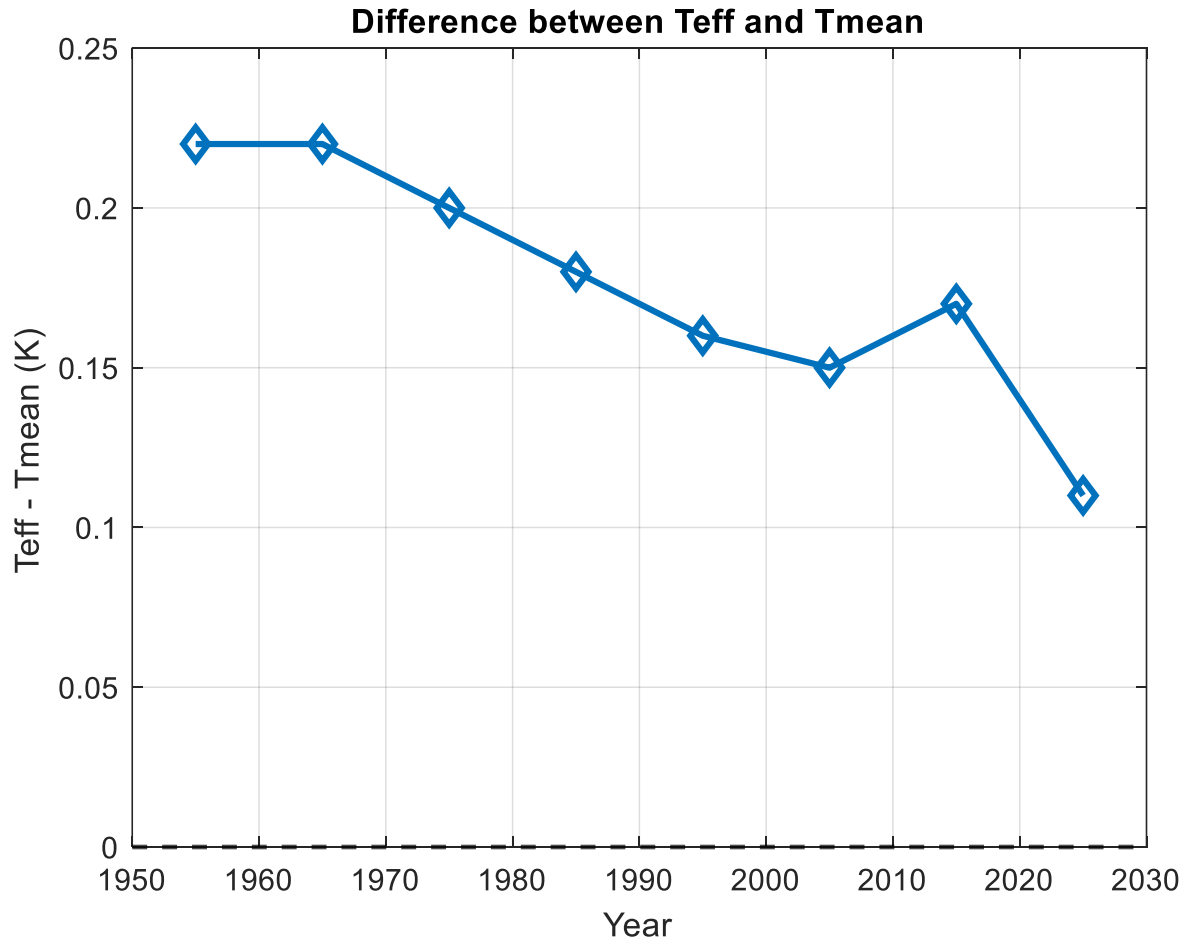

**Figure 2.**

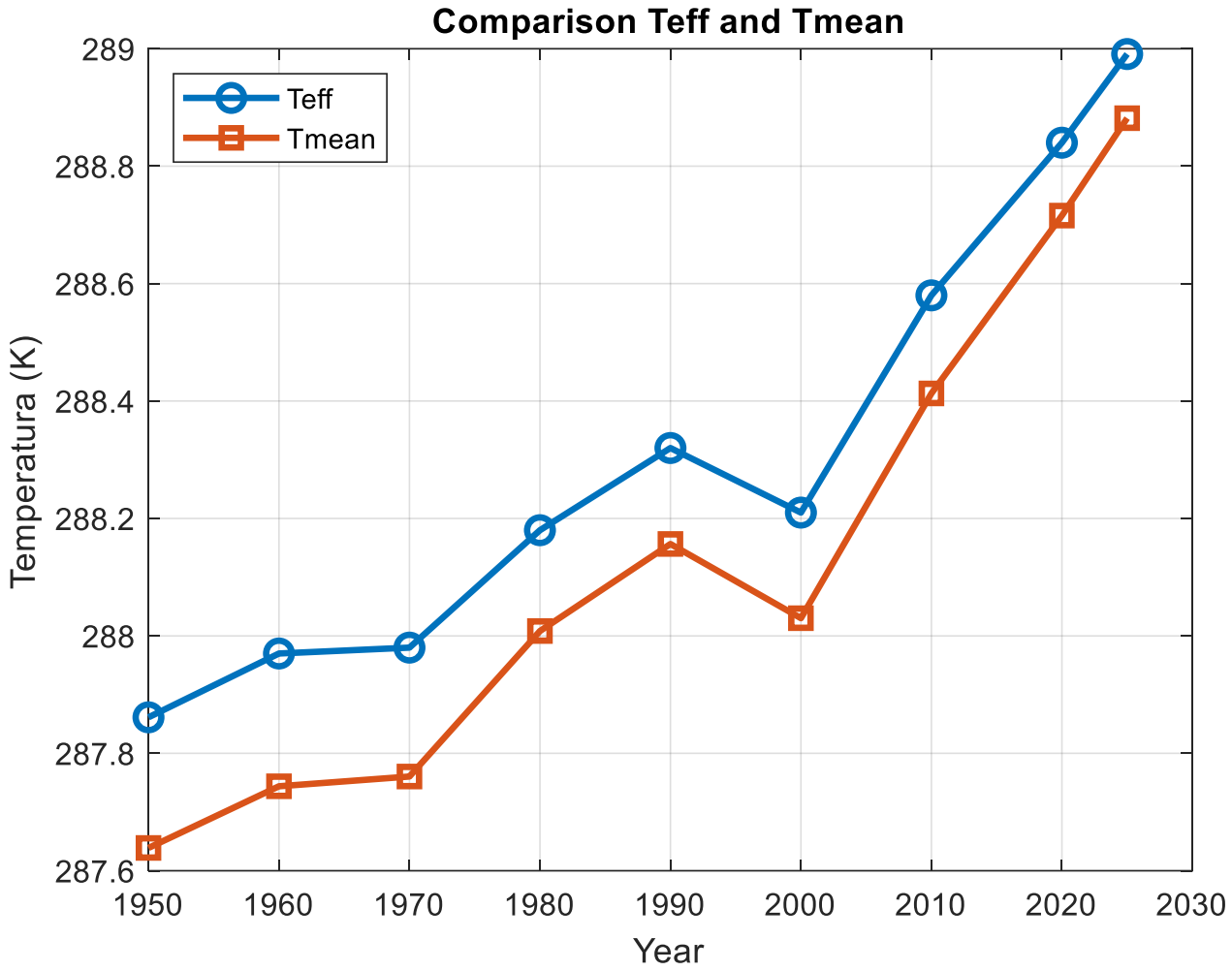

**Figure 3.**

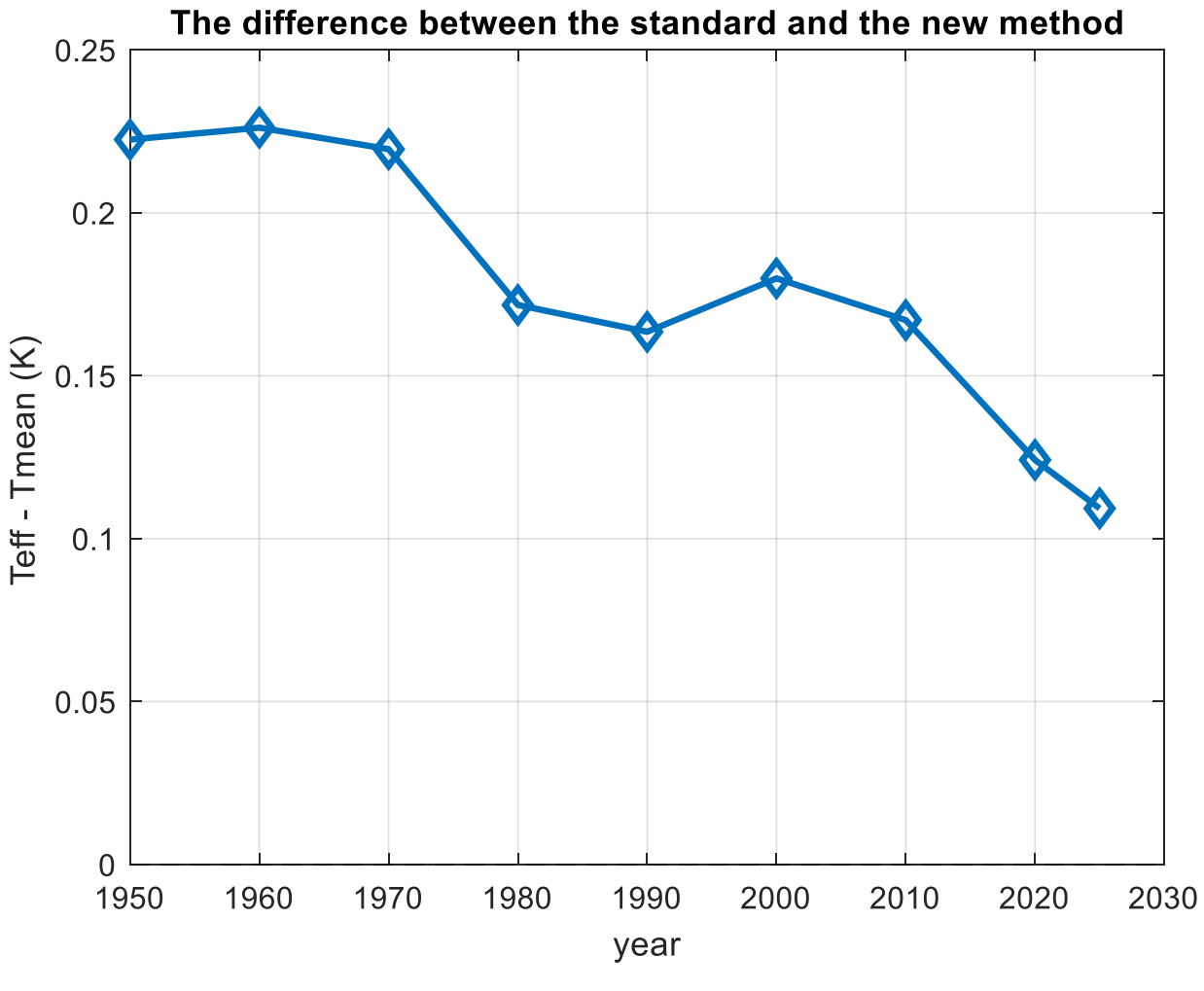

**Figure 4.**

As one can see, there is similar, but not the same, behave of both temperatures with trends of increasing global warming. From the figures 2 and 4, it is clear that the difference between temperatures decreasing. Possible reason for that is that the spatial skin temperature distribution changed during time ie, become more uniform due to surface albedo change **Figure 5.**

This result is not surprising because, although the conventional approach is theoretically inadequate, both methods must yield the similar trend if the spatial temperature distribution remains approximately constant over time. The effective temperatures are always higher than mean temperatures due to non-linearity i.e. in extreme temperature conditions power to forth contribute more **Figure 6.** and **7**.

To check validity of the model for example 2025 year, total outgoing energy calculated by our method is $E_{total} = 6.36 \cdot 10^{24} J$, and derived from *NASA CERES* satellite measurements $6.40 \cdot 10^{24} J$, are compared and this difference is less than $1\%$.

## The Influence of the Atmosphere

Since the analysis is intended for absolute measurements of outgoing energy from the Earth surface, one can now compare right variables and possible causes for changes.

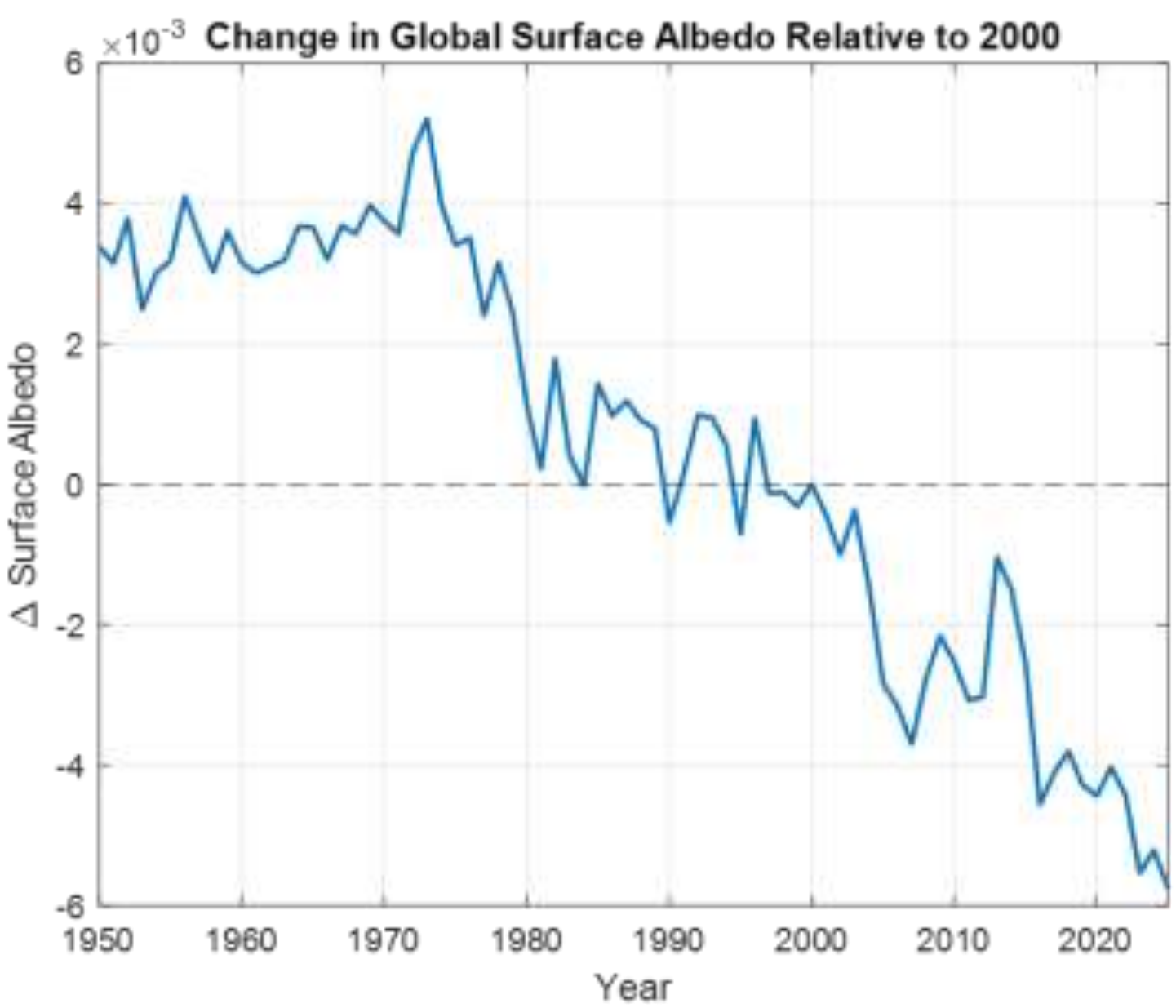

**Figure 5.**

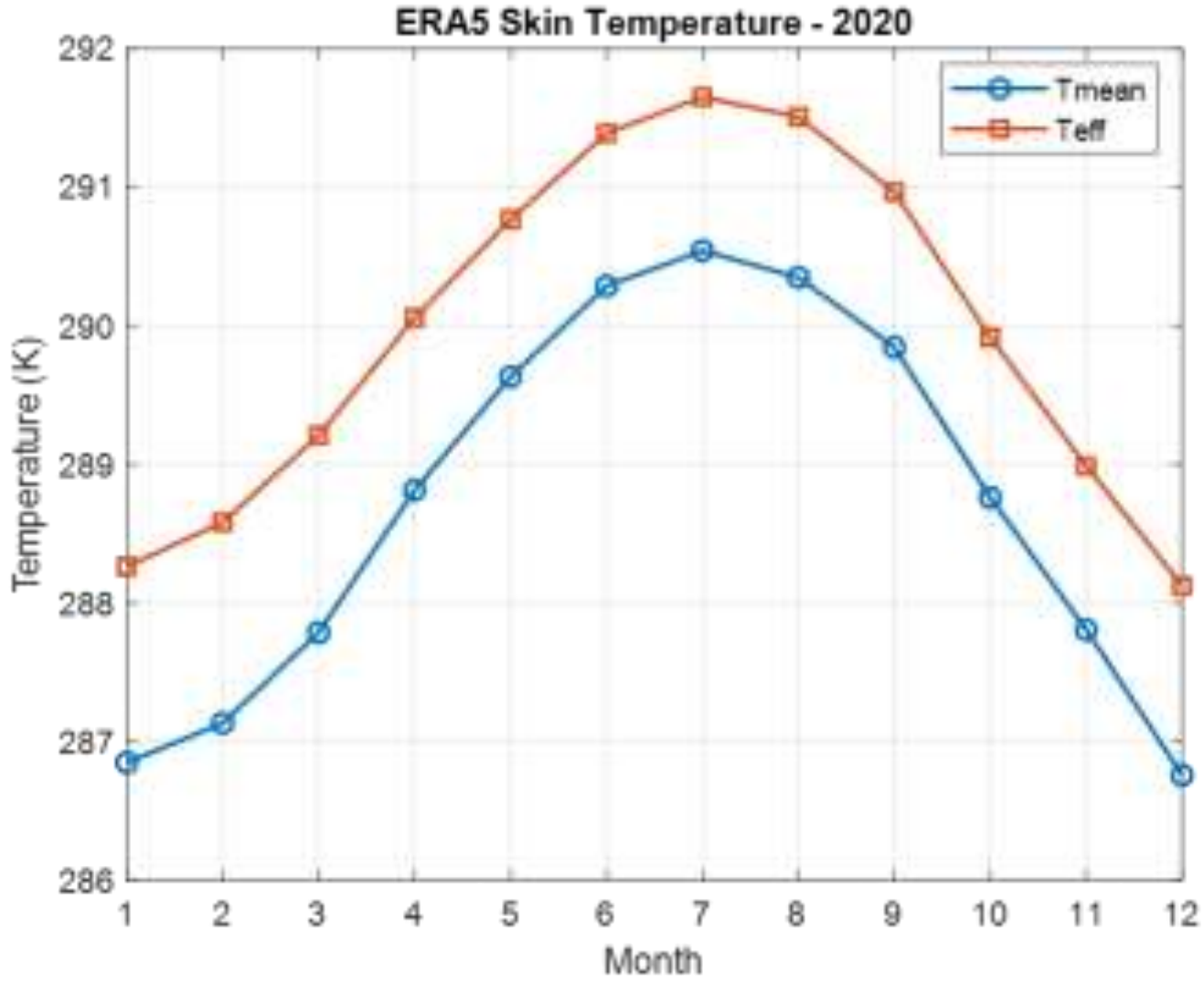

**Figure 6.**

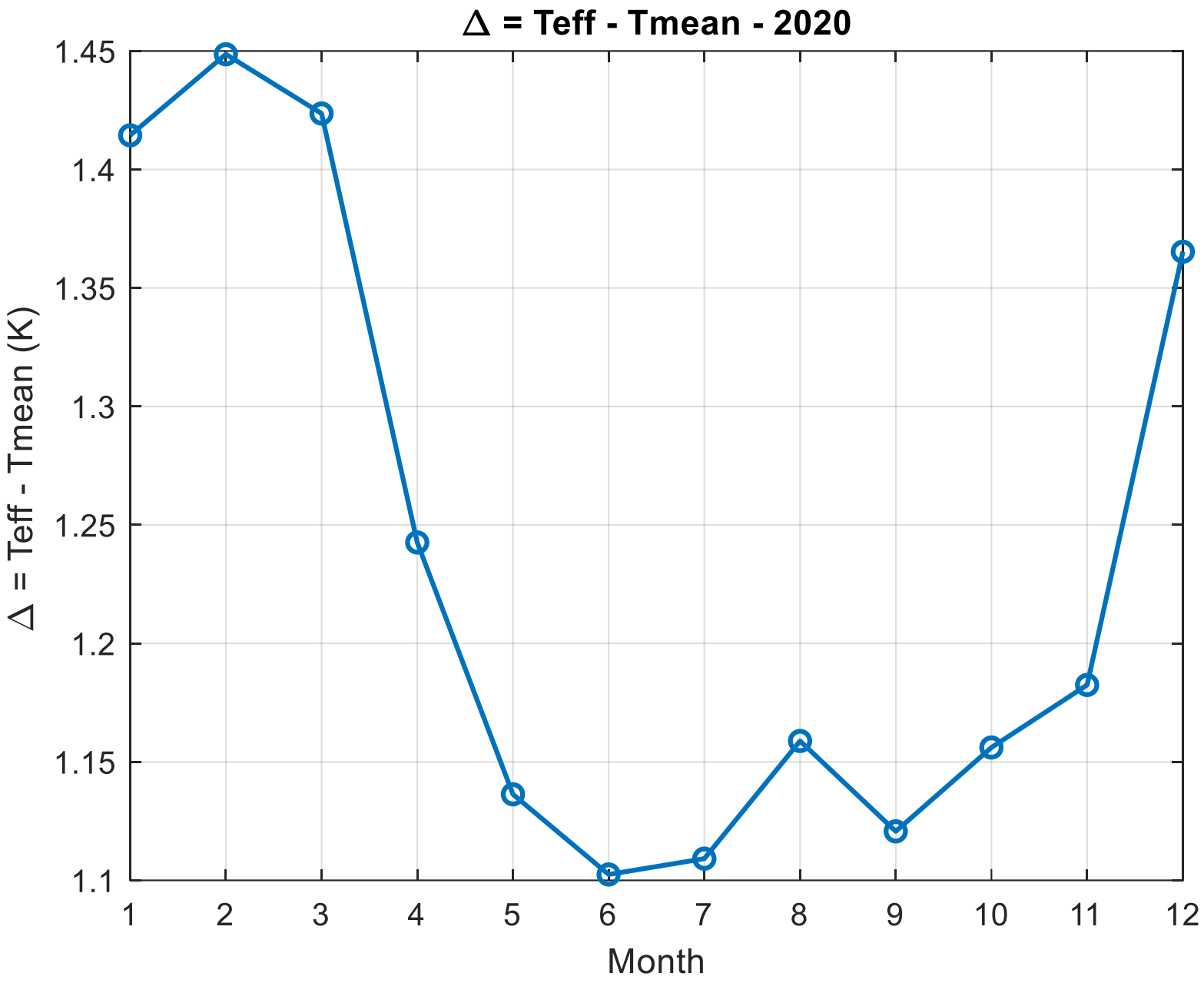

**Figure 7.**

The Earth's temperature is governed by energy balance—solar shortwave radiation heats the planet, while long-wave infrared radiation cools it. The Earth's effective radiometric temperature, calculated as:

$$T_{Earth} = \sqrt[4]{\frac{(1-\alpha)S}{4\sigma}} \approx 255K = -18℃, \tag{3}$$

where S is the solar constant, σ is the Stefan-Boltzmann constant, and α is albedo (~0.3), differs significantly from the actual mean surface temperature (~13.9°C), highlighting the atmosphere's role [5]. So, the statement "'natural' pre-industrial levels of greenhouse gases is 30-35K" is false because compares different variables, the effective and mean temperatures.

Two problems arise immediately, making these comparisons meaningless. First, the albedo differs between a bare planet and a planet with an atmosphere. Second, the temperature calculated using equation (3) assumes a uniformly heated planet, which never actually occurs in reality.

To get more realistic estimation about the atmosphere influence one must estimate new possible values of the Earth's albedo without the atmosphere and put in eq. 3 .

For the albedo value from literature, $\alpha_{Earh\ surface} \approx 0.2$ [6], $T_{eff} = 263K$. If one took albedo like the Moon's $\approx 0.11$, $T_{eff} = 270K$, or the the Earth's current surface albedo $\approx 0.16$, $T_{eff} = 266K$. Thus, in all scenarios these temperatures are much higher than $255K$, that implies less influence of the atmosphere in the Earth's heating.

As one more results is the great negative correlation $r = -0.98$, between effective temperatures and the Earth's surface albedo **Figure 8**, that suggests such possible scenario: A positive albedo feedback is triggered by increasing the temperature (random fluctuation or greenhouse effect), decreasing ice/snow surface, decreasing albedo, increasing temperature etc.

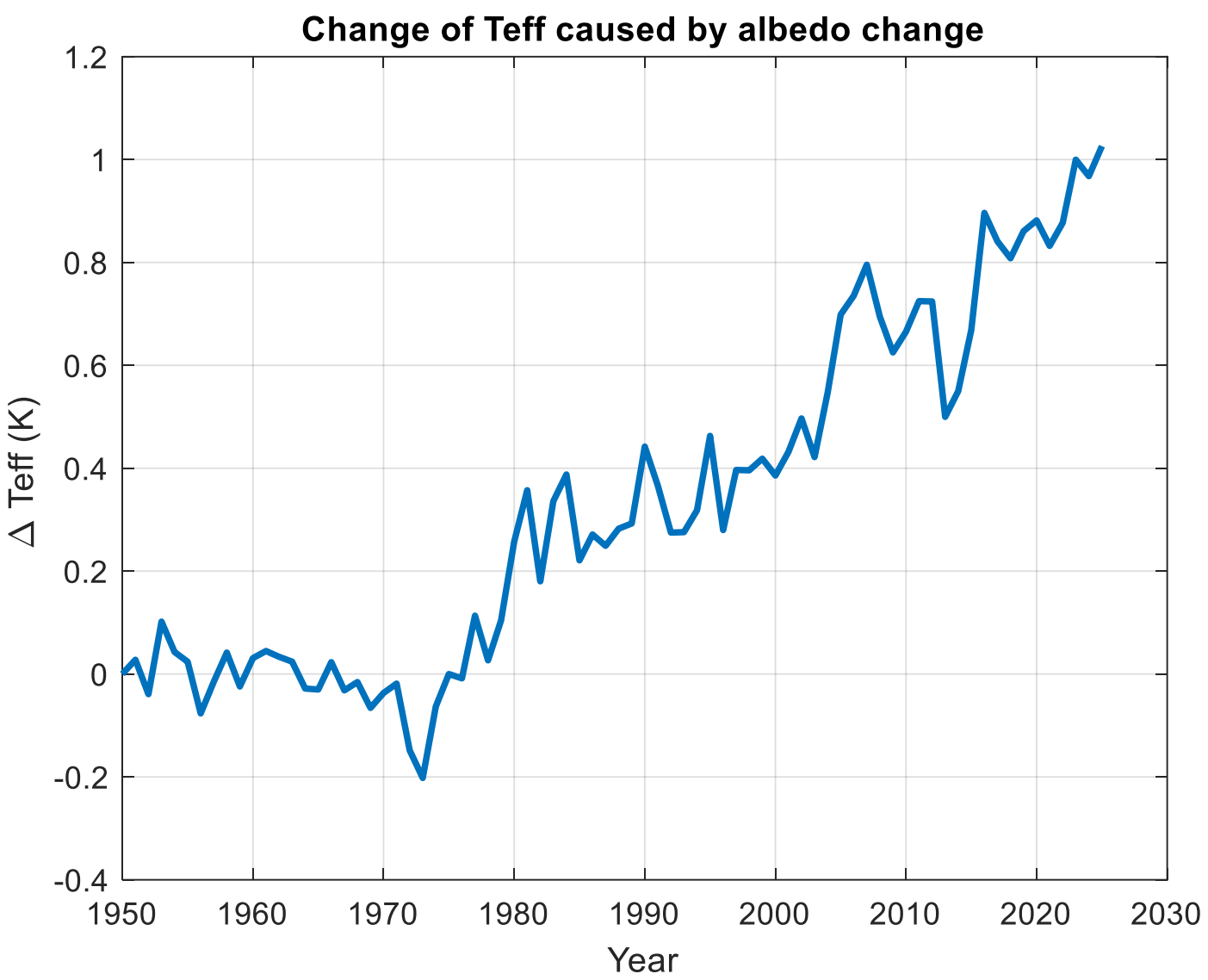

**Figure 8.**

#### CONCLUSION

This leads to the conclusion that the current methodology for estimating global warming and climate change—based on calculating the global mean temperature—is insensitive and inadequate. The primary issue is that the global mean temperature lacks direct physical meaning and has no explicit connection to the energy balance.

Mathematically, this inconsistency arises because the global mean temperature is a linear combination of individual temperatures, whereas the energy budget is highly nonlinear, depending on the fourth power of temperature (as dictated by the Stefan-Boltzmann law).

To address this limitation, we propose a more physically accurate method, which, when tested, provide new insights into climate change using existing datasets. The concept of the "effective temperature for potential cooling" offers a more sensitive indicator of global warming and the impact of increasing greenhouse gas concentrations in the atmosphere. An important advantage of this approach is that it can be easily implemented, as it utilizes existing climate datasets.

By using the new approach is shown that the influence of the atmosphere is highly overestimated.

It is also hypothesized a positive Earth's surface albedo feedback as a main contributor to increasing global warming.

**Acknowledgements**

The author gratefully acknowledges the assistance of artificial intelligence (ChatGPT), which significantly accelerated the implementation of the ideas presented in this work.

## Statements and Declarations

-Funding: *The author declares that no funds, grants, or other support were received during the preparation of this manuscript.*

-Competing interests: *The author declares that has no financial interests.*

-Author contributions: *There is only one author.*

-Ethics approval and consent to participate: *not applicable.*

-Availability of data and materials: *not applicable.*